\documentstyle[12pt,epsf]{article}
\newlength{\egwidth}
\newenvironment{eg}%
{\begin{list}{}{\setlength{\leftmargin}{0.05\textwidth}%
\setlength{\rightmargin}{\leftmargin}}\item[]\normalsize}%
{\end{list}}
\newenvironment{egbox}%
{\begin{minipage}[t]{\egwidth}}%
{\end{minipage}}
\newcommand{\egstart}{\begin{eg}\begin{egbox}}

\newcommand{\egend}{\end{egbox}\end{eg}}

\newcommand{\ltsim}{\;\raisebox{.4ex}{$<$}\hspace{-0.8em}\raisebox{-0.6ex}
                     {$\sim$}\;}
\newcommand{\bea}{\begin{eqnarray}}
\newcommand{\eea}{\end{eqnarray}}
\newcommand{\be}{\begin{equation}}
\newcommand{\ee}{\end{equation}}
\newcommand{\p}{\prime}
\newcommand{\nn}{\nonumber}
\newcommand{\rf}[1]{(\ref{#1})}

\begin{document}

\begin{center}
{\bf NUCLEON STRUCTURE FROM THE QUASITHRESHOLD INVERSE PION ELECTROPRODUCTION}
\bigskip

{\bf T.D. Blokhintseva}\footnote{blokhin@main1.jinr.dubna.su},
{\bf Yu.S. Surovtsev}\footnote{surovcev@thsun1.jinr.ru},\\
{\it Joint Institute for Nuclear Research, 141 980 Dubna, Moscow Region,
Russia}\\
{\bf M. Nagy}\footnote{fyzinami@nic.savba.sk}\\
{\it Institute of Physics, Slov.Acad.Sci., D\'ubravsk\'a cesta 9,
842 28 Bratislava, Slovakia}
\end{center}
\begin{abstract}
The process $\pi N\to e^+e^- N$~(IPE), being a natural and unique
laboratory for studying the hadron electromagnetic structure in the
sub-$N\overline{N}$-threshold time-like region of the virtual-photon ``mass''
$\lambda^2$, turns out to be also very useful for investigating the nucleon
weak structure. A theoretical basis of the methods for extracting practically
model-independent values of the electromagnetic hadron form factors in the
time-like region and for determining the weak structure of the nucleon in the
space-like region from experimental data on IPE at low energies is outlined.
The results of extracting, by those methods, the electromagnetic
$F_1^v(\lambda^2)$ and pseudoscalar $G_P(t)$ form-factors of a nucleon are
presented, where an indication of the existence of the state $\pi^\p$
in the range 500-800 MeV (possibly the first radial excitation of pion) is
obtained, and the coupling constant of this new state with the nucleon is
estimated.
\end{abstract}

\section{Introduction}
Since the processes, responsible for forming the particle structure, run
their course in the region of the time-like momentum transfers, the use of
time-like virtual photons is necessary and promising in studying the form
factors of hadrons and nuclei. Along this way one could rely on obtaining
an interesting (and, maybe, unexpected) information on properties of matter.
For a long time, the process $\pi N\to e^+e^- N$~(inverse pion
electroproduction -- IPE) was a single source of information on the nucleon
electromagnetic structure in the time-like region of the virtual-photon
``mass'' $\lambda^2$. This process is investigated both theoretically
\cite{Geffen}-\cite{Smirn-Shum} and experimentally \cite{Samios}-\cite{Baturin}
from the beginning of 1960's. In Refs.\cite{BST-yaf75,ST-yaf72,ST-yaf75},
we have worked out the method of extracting the pion and nucleon
electromagnetic form factors (FF's) from IPE at low energies. This method has
been successfully realized in experiments on nucleon and
nuclei $^{12}$C and $^7$Li ~\cite{Berezh,Baturin} where first a number of FF
values was obtained in the time-like $\lambda^2$ region from 0.05 to 0.22
(GeV/c)$^2$. In Refs.\cite{Baldin} the use of IPE at intermediate (over $\pi
N$ resonances) energies and small $|t|$ for studying the nucleon
electromagnetic structure is proposed and justified up to $\lambda^2\approx
m_\rho^2$. Though at present the data of measurements of process $p \overline
p \to e^+ e^-$ are available \cite{Bardin}, however, there still remains a
wide enough $\lambda^2$-range (up to $4m^2$) where FF's cannot be measured
directly in those experiments.  On the other hand, at present, with intense
pion beams being available, more detailed experiments are possible aimed both
at extracting the hadron structure and at carrying out the multipole analysis
similar to that for photo- and electroproduction (e.g. \cite{Dev-Lyth-Rank}).
For example, in the $P_{33}(1232)$ region it is interesting to verify the
$\lambda^2$-dependence of the colormagnetic-force contribution, found in the
constituent quark model \cite{Goghilidze}. Therefore, it is worthwhile to
continue the investigation of IPE, to recall the results obtained on the
possibility of studying the electromagnetic and weak structure of a nucleon in
IPE and to give the results of that studying.

A possibility of investigating the nucleon weak structure is based on the
current algebra (CA) description and the remarkable property of IPE according
to which the $e^+e^-$ pairs of maximal masses (at the ``quasithreshold'') are
created by the Born mechanism with the rescattering-effect contributions at
the level of radiative corrections up to the total $\pi N$ energy $w\approx
1500$ MeV (the ``quasithreshold theorem'') \cite{ST-yaf72}. Therefore, the
threshold CA theorems for pion electro- and photoproduction can be justified
in the case of IPE up to the indicated energy \cite{Kulish,Tkeb}. This allows
one to avoid threshold difficulties when using IPE (unlike electroproduction)
for extracting weak FF's of the nucleon. Furthermore, there is no strong
kinematical restriction inherent in $\mu$ capture and is no kinematical
suppression of contributions of the induced pseudoscalar nucleon FF to cross
sections of ``straight'' processes as $\nu N\to lN$ present in due to
multiplying by lepton masses. Information on the pseudoscalar nucleon FF
$G_P$ (which is practically absent for the above reasons) is important because
$G_P$ is contributed by states with the pion quantum numbers and, therefore,
is related to the chiral symmetry breaking.

This paper is organized as follows. In Sect. 2, we outline the method of
determining the nucleon electromagnetic FF's from the low energy IPE and
indicate some results of appllying this method to analyzing IPE data on the
nucleon and nucleus $^7$Li. Section 3 is devoted to extracting the
pseudoscalar nucleon FF from the same IPE data, and an interpretation of the
obtained result is given.

\section{Method of Determining the Nucleon Electromagnetic Structure from
the Low Energy IPE}

For obtaining a reliable information on the nucleon structure, it is important
to find kinematic conditions where the IPE dynamics is determined mainly by
a model-independent part of interactions, the Born one. To this end, we use
such general principles, as analyticity, unitarity, and Lorentz invariance,
and phenomenology of processes $ eN\to e\pi^{\pm} N$ and $\gamma N
\leftrightarrow \pi^{\pm} N$, considered in the framework of the unified
(including IPE) model.

In the one-photon approximation, the amplitude of the process $e N\to e\pi N$
is represented
as
\be \label{1photon-appr.}
\overline{u}(p_2)Tu(p_1)=\frac{m_e^2}{\lambda^2}\varepsilon^\mu
J_\mu(s,t,\lambda^2),
\ee
where
\be \label{lept.current}
\varepsilon^\mu=\overline{u}(k_1)\gamma^\mu u(k_2)
\ee
and
\be \label{hadr.current}
J_\mu=<p_2,q|J_\mu(0)|p_1>
\ee
are matrix elements of lepton and hadron electromagnetic currents,
respectively; $p_1$ and $p_2$ are momenta of nucleons, $k_1,k_2$ and $q$ are
momenta of electrons and pion, furthermore,
$$(\widehat{p}-m)u(p)=0,~~~~\widehat{p}=p^\mu\gamma_\mu,~~~~~~~
p_1^2=p_2^2=m^2,~~~k_1^2=k_2^2=m_e^2,~~~q^2=m_\pi^2; $$
$s=(p_1+q)^2, ~t=(k-q)^2$~ are usual Mandelstam variables ($k=k_1-k_2$ is
the momentum of a virtual photon $\gamma^*$, which is space-like in
electroproduction, $k^2=\lambda^2<0$). From conservation of lepton and hadron
electromagnetic currents it follows that $J_\mu k^\mu=
\varepsilon_\mu k^\mu=0$.

Under the assumption of $T$-invariance, for the IPE process, one must only take
the spinor $v(k_2)$ instead of $u(k_2)$ in the lepton current
\rf{lept.current}. Then the $\gamma^*$-quantum momentum $k=k_1+k_2$ is
time-like, and ~$4m_e^2\leq\lambda^2\leq(\sqrt{s}-m)^2$~ is the range of
$\lambda^2$ values for given $s$.

So, the research of pion photo-, electroproduction and IPE in the one-photon
approximation is related with studying the amplitude of the process
~$\gamma^* N\leftrightarrow\pi N^\p$, ~$J_\mu(s,t,\lambda^2)$, ~where
~$\lambda^2=0,~<0$~ and ~$>0$~ correspond to the above three processes,
respectively. That approach permits us to predict peculiarities of the IPE
dynamics on the basis of a rich experimental material on electro- and
photoproduction for testing a reliability of the unified model of these three
processes.

Further, the current $J_\mu(s,t,\lambda^2)$ can be expanded in six
independent covariant gauge-invariant structures $M_i$ \cite{BST-yaf75,Adler}:
\be \label{J:covar.expantion}
J_\mu\varepsilon^\mu=\sum_{i=1}^6 A_i(s,t,\lambda^2)
\overline{u}(p_2)M_iu(p_1),
\ee
where
\bea \label{M_i}
\left.\begin{array}{ll}
M_1=\frac{i}{2}\gamma_5\gamma^\mu\gamma^\nu F_{\mu\nu}, &~~M_4=
2i\gamma_5\gamma^\mu P^\nu F_{\mu\nu}-2mM_1,\nn \\
M_2=-2i\gamma_5 P^\mu q^\nu F_{\mu\nu},
&~~M_5=-i\gamma_5 k^\mu q^\nu F_{\mu\nu},\nn \\
M_3=i\gamma_5 \gamma^\mu q^\nu F_{\mu\nu}, &~~
M_6=i\gamma_5 k^\mu\gamma^\nu F_{\mu\nu}\nn
\end{array}  \right.
\eea
with $F_{\mu\nu}=\varepsilon_\mu k_\nu-k_\mu\varepsilon_\nu$ and
$P=\frac{1}{2}(p_1+p_2)$; $A_i(s,t,\lambda^2)~(i=1,\cdots,6)$ are independent
invariant amplitudes, free from kinematic constraints, but $A_2$ and $A_5$
have a kinematic pole at $t=m_\pi^2+\lambda^2$. For real photoproduction the
amplitudes $A_5$ and $A_6$ are absent.

When constructing a dynamic model in the first resonance region, we take
into account the experimental fact that $P_{33}(1232)$ resonance is mainly
excited by the isovector magnetic component of photon in photo- and
electroproduction. Now, let us notice that, in accordance with the
conventional procedure of Reggeization, one can obtain that as $s\to\infty$
and at small $|t|$ invariant amplitudes behave as \cite{Sur}
$$A_i\sim s^{\alpha(t)-1}~(i\neq 5),~~~~~~~~~~~~~~~ A_5\sim s^{\alpha(t)}.$$
Therefore, in a complete $s$-channel description, with taking
crossing-properties of the amplitudes $A_i$ into account, we should write a
fixed-$t$ dispersion relation with one subtraction at a finite energy for
the isovector amplitude $A_5^{(-)}$; and without subtractions, for remaining
amplitudes. However, the dispersion integrals with spectral functions
describing the magnetic excitation of the P$_{33}$(1232) resonance converge
very well already at $\sim$ 2 GeV for all the amplitudes $A_i$. Therefore,
for isovector amplitudes $A_i^{(\pm)}$ we shall use fixed-$t$ dispersion
relations without subtraction at the finite energy \cite{BST-yaf75,Adler}
\bea \label{eq:DR}
& & A_i^{(\pm)}(s,t,\lambda^2)=\tilde{R}_5^{(-)}+c_5+R_i^{(\pm)}\Bigl(\frac{1}
{m^2-s}\pm\frac{\epsilon_i}{m^2-u}\Bigr)+\nn\\
& &~~~~~~+\frac{1}{\pi}\int_{(m+m_\pi)^2}^{\infty} ds^\p~ \mbox{Im}
A_i^{(\pm)}(s^\p,
t,\lambda^2)\Bigl(\frac{1}{s^\p-s-i\varepsilon}\pm\frac{\epsilon_i}{s^\p-u}
\Bigr),
\eea
and for isoscalar amplitude we take the Born approximation
\be \label{eq:A(0)}
A_i^{(0)}(s,t,\lambda^2)~=~R_i^{(0)}\Bigl(\frac{1}{m^2-s}+\frac{\epsilon_i}
{m^2-u}\Bigr),
\ee
where ~~$\epsilon_{1,2,4}=-\epsilon_{3,5,6}=1,\quad u=2m^2+m_\pi^2+\lambda^2-
s-t$,
\bea \label{eq:Born}
&& R_1^{(\pm,0)}=-\frac{g}{2} F_1^{v,s} (\lambda^2),\qquad
~~~~~~~~R_2^{(\pm,0)}=\frac {gF_1^{v,s} (\lambda^2)}{t-m_\pi^2-\lambda^2},\nn\\
&& R_3^{(\pm,0)}=R_4^{(\pm,0)}=(-,+)\frac{g}{2} F_2^{v,s} (\lambda^2), \qquad
~R_5^{(\pm,0)}=R_6^{(\pm,0)}=0,\\ &&
\tilde{R}_5^{(-)}=\frac{2g}{\lambda^2}~
\biggl[\frac{F_1^v(\lambda^2)}{t-m_\pi^2-\lambda^2}-\frac{F_\pi(\lambda^2)}
{t-m_\pi^2}\biggr] \nn
\eea
with the $\pi N$ coupling constant $g^2/4\pi=14.6$ and with the following
normalization of the form factors: $~~ F_1^{v,s}(0)=F_\pi(0)=1,
~~2m F_2^v(0)=3.7,~~ 2m F_2^s(0)=-0.12$ and
\be \label{eq:c5}
c_5=\frac{2}{m_\pi^2+\lambda^2-t}~\frac{1}{\pi}\int_{(m+m_\pi)^2}^{\infty}
\frac{ds^\p}{s^\p-m^2}~\lim_{t \to {m_\pi^2+\lambda^2}}\bigl[(t-m_\pi^2-
\lambda^2)\mbox{Im} A_5^{(-)}(s^\p,t,\lambda^2)\bigr].
\ee
The terms $\tilde{R}_5^{(-)}$ and $c_5$ belong only to the amplitude
$A_5^{(-)}$. Note that although $A_2$ and $A_5$ have a kinematic pole at
$t=m_\pi^2+\lambda^2$, these amplitudes enter into the matrix element through
the combination ~$(s-m^2)A_2+\lambda^2 A_5$, which, in turn, is equal to
~$2B_3-B_2$~($B_2$ and $B_3$, Ball's amplitudes, have been proved to have
no kinematic singularities \cite{Adler}), therefore, this singularity is
cancelled out kinematically. However, in specific model calculations, a
singularity at $t=m_\pi^2+\lambda^2$ being absent is guaranteed by the
condition
\be \label{eq:Adler-cond}
\lim_{t \to {m_\pi^2+\lambda^2}}(t-m_\pi^2-\lambda^2)\bigl[(s-m^2)A_2+
\lambda^2 A_5\bigr]=0,
\ee
The term $c_5$ ensures \rf{eq:Adler-cond} to be valid \cite{Adler}.

For the spectral functions $\mbox{Im} A_i^{(\pm)}(s^\p,t,\lambda^2)$ we
suppose that they are defined by the magnetic excitation of the
$P_{33}(1232)$ resonance :
\be \label{eq:ImA}
\mbox{Im} A_i^{(\pm)}(s,t,\lambda^2)=\frac{4\pi}{3}{ 2 \choose {-1} }\frac
{G_M^v(\lambda^2)\sin^2 \delta_{33}(w)}{gm_\pi q^3 [(w+m)^2-\lambda^2]}~
a_i(w,t,\lambda^2),
\ee
where \qquad $w=\sqrt {s}, \qquad G_M^v=F_1^v+2mF_2^v$,~ $\delta_{33}(w)$ is
the corresponding phase-shift of the $\pi N$-scattering amplitude, and
\bea \label{eq:ai}
&& a_i(w,t,\lambda^2)=\alpha_i(w,t)-\lambda^2 \beta_i(w),\qquad
a_{2,5}(w,t,\lambda^2)=\frac{\alpha_{2,5}(w,t)-\lambda^2 \beta_{2,5}(w)}
{t-m_\pi^2-\lambda^2},~~~~\\
&&~~~~~~~i=1,3,4,6 \nn
\eea
and the coefficients $\alpha_i,\beta_i$ have the form
\bea \label{coef.in-ai}
\left.\begin{array}{ll}
\alpha_1=\frac{1}{2}(w+m)[(w+m)q_0-m_\pi^2+3t], &~~\beta_1=
\frac{1}{2}(w+m+q_0),           \\
\alpha_2=\frac{3}{2}(w+m)(m_\pi^2-t), &~~\beta_2=\frac{1}{2}(w+m)+q_0, \\
\alpha_3=-\frac{1}{2}(w+m)(w+m-q_0)-\frac{3}{4}(m_\pi^2-t), &~~
\beta_3=-\frac{3}{4}, \\
\alpha_4=(w+m)(w+m+\frac{1}{2}q_0)-\frac{3}{4}(m_\pi^2-t), &~~
\beta_4=\frac{3}{4},\\
\alpha_5=2(s-m^2)(w+m+\frac{1}{2}q_0)-\frac{3}{2}(w-m)(m_\pi^2-t), &~~
\beta_5=\frac{3}{2}(w-m),\\
\alpha_6=-\frac{1}{2}(w+m)q_0)-\frac{1}{4}(m_\pi^2+3t), &~~
\beta_3=-\frac{3}{4}.
\end{array}  \right.
\eea
Furthermore, according to the results of the photoproduction
multipole analyses \cite{Dev-Lyth-Rank}, we take $E_{0+}^{(0)}=0$ above the
P$_{33}$(1232) energy.

Note that it is a first reliable version of the model for a unified treatment
of contemporary experimental data on pion photo-, electroproduction and IPE
in the energy region from the threshold up to the second $\pi N$ resonance.
A more subtle model requires to consider, in addition to the isovector
quadrupole excitation of the $P_{33}$(1232) resonance ($E_{1+}^{\pm}$ and
$L_{1+}^{\pm}$), the isoscalar excitation of this resonance and contributions
of other $\pi N$ isobars and high-energy ``tails'' to the absorption parts of
amplitudes -- for a balanced account of small corrections to the main version
of the model. Furthermore, notice that for $k^2>m_\pi^2$, in dispersion
integrals there is an unobservable region
$(m+m_\pi)^2\leq s\leq (m+\lambda)^2$, the analytic continuation into which
of the approximation \rf{eq:ImA} is immediate. However, for the analytic
continuation into this region of the corrected absorption parts of amplitudes,
expanded in eigenfunctions of the angular momentum, one must use
quasithreshold relations (following from causality-analyticity) between
electric and longitudinal multipoles \cite{Sur}, where ``toroid'' multipoles
play up. However, at a level of contemporary experimental data, the
above-stated model is sufficient.

Earlier it was shown that the model, based on the fixed-$t$ dispersion
relations without subtractions at a finite energy for isovector amplitudes
with the spectral functions describing the magnetic excitation of the
P$_{33}$(1232) resonance and with the isoscalar amplitudes being the Born ones
\cite{Adler}, is successful in the unified explanation of the experimental
data on pion electro-, photoproduction and IPE in the total-energy region from
the threshold up to $w\approx 1500$ MeV \cite{BST-yaf75}.

Application of this model to the calculations for IPE shows the interesting
growth of the relative contribution of the Born terms with $\lambda^2$
\cite{BST-yaf75}. This approximate dominance of the Born terms has a
model-independent explanation and is related with the quasithreshold
theorem \cite{ST-yaf72} which means that at the quasithreshold
($|{\vec k}|\to 0, ~\lambda^2\to\lambda_{max}^2=(\sqrt{s}-m)^2$) the IPE
amplitude becomes the Born one in the energy region from the threshold up to
$\sim 1500$ MeV. That remarkable dynamics of IPE distinguishes it essentially
from photo- and electroproduction, where rescattering effects are
$\sim 40-50\%$. Let us explain the quasithreshold behaviour of the IPE
amplitude. At $|{\vec k}|\to 0$ multipole amplitudes behave as
\bea \label{q.thr.mult.ampl.}
&& M_{l\pm}\propto k^l,~~~~~~~E_{l+}\propto k^l,~~~~~~~L_{l+}\propto k^l,\nn\\
&& E_{l-}\propto k^{l-2},~~~~~~~L_{l-}\propto k^{l-2},
\eea
therefore, at $|{\vec k}|= 0$ only the electric ($E_{0+}$ and $E_{2-}$) and
longitudinal ($L_{0+}$ and $L_{2-}$) dipoles survive, and the selection rules
appear (from parity conservation and from that the stopped virtual photon has
the angular momentum $J=1$): at the
quasithreshold only the resonances with $J^P={\frac{1}{2}}^-~(S_{11}(1535),
S_{31}(1650),S_{11}(1700)$, etc.) and $J^P={\frac{3}{2}}^-~(D_{13}(1520),
D_{33}(1670)$,etc.) survive in the $s$-channel of IPE. Furthermore, indeed, in
this kinematic configuration the process is stipulated only by two independent
dipole transitions (either electric or longitudinal), because from the
causality (analyticity) the quasithreshold constraints arise:
\be \label{q-thr-constr} E_{0+}=L_{0+}, \qquad ~~~~~~E_{2-}=-L_{2-}.  \ee
Since the $s$- and $d$-wave $\pi N$ resonances are excited above 1500 GeV, one
can expect that dipoles $E_{0+}$ and $E_{2-}$ are mainly the Born ones below
this energy. All the multipole analyses of charged pion photoproduction agree
with this; and e.g., the dispersion-relation calculation has confirmed this
fact at $\lambda^2\neq 0$. Therefore, with a good accuracy ($<5\%$), the
quasithreshold IPE amplitude is the Born one in this region, and we can
write for the quasithreshold IPE below $\sim 1500$ GeV:
\bea \label{q-thr-sigma}
\lim_{k\to 0}~\frac{q}{k}~\frac{d^2\sigma}{d\lambda^2d\cos\theta} & \approx &
\frac{\alpha}{12\pi}~\frac{m^2}{(\sqrt{s}-m)^2}~\frac{1}{s}~\bigl\{(1+\cos^2
\theta)|E_{0+}^{Born}+E_{2-}^{Born}|^2+\nn\\ & &
~~~~~~~~~~~~~~~~~~~~~~~~~~~~+\sin^2\theta|E_{0+}^{Born}-2E_{2-}^{Born}|^2\bigr\}.
\eea
Here $\theta$ is the angle betweem momenta of a final nucleon and of an
electron in the $e^+e^-$ c.m. system.

In fact, in real experimental conditions one is forced to move away from the
quasithreshold, therefore, the realistic model (presented above) is needed.
The so-called ``compensation curves''\cite{ST-yaf75} should help to choose the
optimal geometry of experiment for deriving form factors, these curves being
the ones in the $(s,t)$ plane along which the differential cross-section is
the Born one. These curves are constructed on the basis of comparison of
photoproduction experimental data with the Born cross-section using the
existence theorem for implicit functions.

The method of determining electromagnetic FF's from low energy IPE is based on
using the quasithreshold theorem, the realistic (dispersion relation) model
and the compensation curve. This method has successfully been realized in
experiments on nucleon and on nuclei $^{12}$C and $^7$Li \cite{Berezh,Baturin}
where first a number of FF values was obtained in the time-like $\lambda^2$
region from 0.05 to 0.22 (GeV/c)$^2$.
\begin{table}[htb] \begin{center} Table 1.
\end{center} \begin{center} \begin{tabular}{|l|cccccccccc|} \hline
$\stackrel{}{\lambda^2}, ~m_\pi^2$ & 2.77 & 2.98 & 3.44 & 3.75 & 4.00 & 4.47 & 4.52 & 5.28 & 5.75 & 6.11 \\
\hline
$\stackrel{}{F_1^v(\lambda^2)}$ & 0.96 & 0.93 & 1.16 & 1.04 & 1.14 & 1.22 & 1.13 & 1.20 & 1.32 & 1.36 \\
\hline
$\stackrel{}{F^\pi(\lambda^2)}$ & 0.91 & 0.85 & 1.04 & 0.91 & 0.99 & 1.04 & 0.95 & 1.01 & 1.12 & 1.16 \\
\hline
$\stackrel{}{\rm Error}$ & 0.10 & 0.09 & 0.10 & 0.08 & 0.16 & 0.10 & 0.09 & 0.09 & 0.10 & 0.08 \\
\hline
\end{tabular}
\end{center}
\end{table}
In Table 1, the values of electromagnetic FF's, obtained in experiments on
nucleons, we need later, are presented. Note that here the same experimental
errors are cited for $F_1^v$ and $F^\pi$ because in this $\lambda^2$-range
these FF's can be considered to be connected with each other by the relation
$F_1^v(\lambda^2)-F^\pi(\lambda^2)=\bigtriangleup(\lambda^2)$. The quantity
$\bigtriangleup(\lambda^2)$ has been taken from the dispersion calculations
\cite{Hohler}, and its theoretical uncertainty is significantly less than the
one in the calculations of $F_1^v$ and $F^\pi$ in view of the compensation of
a number of contributions to the spectral functions and due to the dominating
influence of the contribution of the one-nucleon exchange in this quantity in
the region $4m_\pi^2\leq\lambda^2\ltsim 20m_\pi^2$.

Here we outline also the results of an analysis of the experiment on IPE on
$^7$Li nucleus with $\pi^+$ beam at 500~MeV/c \cite{Baturin}. The missing
mass analysis of the data has shown that about a half of events belongs to the
process\\
\hspace*{4cm}(I)~~~~~~~~~~$\pi^+ + \;^7\mbox{Li}\; \rightarrow e^+ e^-
+ \;^7\mbox{Be}\;$.\\
The remaining events are related to disintegration processes of a nucleus
which are dominated by the reaction\\
\hspace*{4cm}(II)~~~~~~~~~$\pi^+ + \;^7\mbox{Li}\; \rightarrow e^+ e^-
+ p + \;^6\mbox{Li}\;$.\\
When analyzing all the events (with and without disintegration of the nucleus),
the cross section on a nucleus has been supposed to be additively connected
with the cross section on an individual nucleon (taking screening into
account) \cite{Baturin}. The following results for $F_1^v(\lambda^2)$ are
obtained:
\begin{table}[htb]
\begin{center} Table 2. \end{center}
\begin{center}
\begin{tabular}{|c|c|c|c|} \hline
$\stackrel{}{\lambda^2}, ~{\rm GeV}^2$ & 0.09 & 0.15 & 0.22 \\ \hline
$\stackrel{}{F_1^v(\lambda^2)}$ & $1.60 \, \pm \, 0.21$ & $1.53 \, \pm \, 0.09$
& $1.88 \, \pm \, 0.10$ \\
\hline
\end{tabular} \end{center}
\end{table}

In analysing reaction II one assumed that the pion-nucleus amplitude is
determined by the neutron-pole mechanism. The corresponding cross section is
written in the form:
\be \label{sigma-Li-Li}
\sigma_{theor} = A~\left|\frac{2\mu G(Q^2)}{Q^2+2\mu\varepsilon}\right|^2
\sigma(\pi^+ n\to e^+ e^- p)
\ee
where $A$ is a kinematic factor, $G(Q^2)$ is the vertex function of the
$\;^7\mbox{Li}\; \to \;^6\mbox{Li}\;+n$ process (calculated in the
nucleon cluster model \cite{Avakov}), $\mu=\frac{6}{7}m$ and Q are the reduced
mass and the relative 3-momentum of neutron and $^6$Li, $\varepsilon$ is the
$^7$Li binding energy with respect to decay into $^6$Li and $n$. Then we
obtain
\be \label{F1-Li-Li}
F_1^v (0.14 {\rm GeV}^2) = 1.51 \, \pm \, 0.12 .
\ee

The obtained $F_1^v$ values are quite consistent with the calculations in the
framework of the unitary and analytic vector-meson dominance model of the
nucleon electromagnetic structure \cite{Dubnicka}.

\section{Pseudoscalar Form Factor of Nucleon from the quasithreshold IPE}

Now, let us indicate another interesting possibility of investigating the weak
nucleon structure related to the nucleon Gamov -- Teller transition described
by the matrix element
\be \label{Gamov-Teller}
\left<N(p_2)|A_\mu^\alpha|N(p_1)\right>=\overline{u}(p_2)\frac{\tau^\alpha}{2}
\Bigl[\gamma_\mu G_A(k^2)+k_\mu G_P(k^2)\Bigr]\gamma_5u(p_1),
\ee
where $A_\mu^\alpha$ is the axial-vector current, $G_A(k^2)$ and $G_P(k^2)$
are the axial and induced pseudoscalar FF's.

An alternative description of IPE in the framework of the current commutators,
PCAC and completeness allows one to derive a low-energy theorem at the
threshold (${\vec q}=0, \lambda^2\rightarrow m_\pi^2$) related to approximate
chiral symmetry and $O(m_\pi^2)$ corrections; and the quasithreshold
minimization of the continuum contribution makes it possible to justify this
approach up to $w\approx 1500$ MeV \cite{Kulish} with the continuum
corrections being practically the same as in the dispersion-relation
description.
Then, at the quasithreshold, retaining only the leading terms in
$\lambda^2/m^2, t/m^2,$ one obtains for the longitudinal part of the
$\pi^- p \rightarrow \gamma^* n$ amplitude (Furlan G. et al. in
ref.\cite{Kulish})
\bea \label{long.part:CA}
E_{0+}-2E_{2-}&=&\frac{\lambda}{2m_\pi^2 f_\pi}
\sqrt{\frac{(w+m)^2-m_\pi^2}{mw}}\Bigl\{D(t)-\Bigl(1+\frac{\lambda}{2m}\Bigr)
D(m_\pi^2-\lambda^2) +\nn\\&&+\frac{m_\pi^2}{2m}\Bigl[G_A(m_\pi^2-\lambda^2)-
\frac{t}{2m}G_P(m_\pi^2-\lambda^2)\Bigr]\Bigr\},
\eea
where the constant of the $\pi\rightarrow\mu+\nu_\mu$ decay $f_\pi$ is
defined by ~$\left<0|A_\mu(0)|\pi(q)\right>=if_\pi q_\mu$,
$D(t)=-2mG_A(t)+tG_P(t),$ and the quasithreshold values of the variables are
$$w_{q.thr.}=m+\lambda, \quad t_{q.thr.}=(m_\pi^2-\lambda^2)\frac{m}
{m+\lambda}.$$

$G_A$ has been measured in various experiments (first of all, in
$\nu n\rightarrow {\mu}^-p,\bar\nu p\rightarrow {\mu}^+n$). It is reasonable
to use first this result:
\be \label{G_A}
G_A(t)=G_A(0)\Bigl(1-t/M_A^2\Bigr)^{-2}, \qquad G_A(0)=-1.25,\quad
M_A=(0.96\pm 0.03) {\rm GeV}.
\ee
However, $G_P$ can be seen to be kinematically suppressed in these experiments
in view of its contribution to cross sections to be multiplied by lepton
masses (from here, a difficulty of obtaining informarion on $G_P$ in these
experiments). In the $\mu$-capture and $\beta$-decay experiments, there is
a strong kinematical restriction of the range $|t|\sim 0-0.01
({\rm GeV}/{\rm c})^2$ in which the weak FF's can be determined, however, with
a large error. For example, its measured value for $\mu$-capture in hydrogen
\cite{Bard} is ~~$G_P(-0.88m_\mu^2 )=-8.7\pm1.9$.~ Recently, $G_P$ has been
measured in the capture of polarized muons by $^{28}$Si nuclei
\cite{Brudanin}.

From formula \rf{long.part:CA} it is seen that the kinematic suppression of
$G_P$ would be absent when the IPE data at the quasithreshold are used for
extracting $G_P$. On the basis of this method, $G_P(t)$ could be determined in
the range up to $t\approx -15m_\pi^2$~ (which corresponds to $w\approx 1500$
MeV). Due to working at the quasithreshold, one succeeds in avoiding threshold
difficulties which are the case when using the analogous method for analyzing
electroproduction data.

Further we shall follow the method of work \cite{Tkeb}. First, using the
$F_1^v(\lambda^2)$ and $F_\pi(\lambda^2)$ values obtained in the analysis
of the IPE data on the nucleon \cite{Berezh} we obtain ten points (which can
be considered as the experimental ones) for the longitudinal part of the
$\pi^-p \rightarrow \gamma^* n$ amplitude at the quasithreshold (Fig.1).
For $G_P(t)$ we take the following dispersion relation without subtractions
\be \label{G_P:DR}
G_P(t)=\frac{2f_\pi g_{\pi N}}{m_\pi^2-t}+\frac{1}{\pi}\int_{9m_\pi^2}^
{\infty} \frac{\rho(t^\p)}{t^\p-t}dt^\p.
\ee
The residue in the pole $t=m_\pi^2$ is determined by the PCAC relation. In
Fig.2, possible contributions to $G_P$ are depictet. When only the $\pi$-pole
term is considered, it is inconsistent with experimental data (the dashed
curve in Fig.1). Since the contributions of non-resonance three-particle
states must be suppressed by the phase volume, it is reasonable to approximate
the integral in \rf{G_P:DR} by a pole term. A satisfactory description is
obtained if
\be \label{G_P}
G_P(t)=G_P^\pi(t)-\frac{2f_{\pi^\p} g_{\pi^\p N}}{m_{\pi^\p}^2-t},\quad
2f_{\pi^\p} g_{\pi^\p N}=(1.97\pm 0.18){\rm GeV},\quad m_{\pi^\p}=0.5 {\rm
GeV},
\ee
where ~~ $G_P^\pi(t)=2f_\pi g_{\pi N}/(m_\pi^2-t)$, the $\pi^\p$ weak-decay
constant $f_{\pi^\p}$ is defined by ~$\left<0|A_\mu(0)|\pi^\p(q^\p)\right>=
if_{\pi^\p} q_\mu^\p$, $g_{\pi N} (=13.5)$ and $g_{\pi^\p N}$ are the coupling
constants of the $\pi$ and $\pi^\p$ states with the nucleon. As it is seen
from the definitions of the weak-decay constants, one must expect that
~$f_{\pi^\p}\ll f_\pi$, to reflect a tendency of another way (in addition to
the Goldstone one) in which the axial current is conserved for vanishing
quark masses. That behaviour is demonstrated in various models with some
non-locality which describe chiral symmetry breaking \cite{Volkov,Kalin}. Note
that the pole at $t={m_{\pi^\p}}^2$ in eq.\rf{G_P}, situated considerably
lower than the poles of the known contributing states $\pi^\p(1300)$ and
$\pi^\p(1770)$, is highly required for describing the obtained experimental
data on IPE.

In Fig.3, the ratio $G_P(t)/G_P^\pi(t)$ is shown. One can see that $G_P(t)$ is
determined by this method with a high accuracy. For the comparison, the $G_P$
values, obtained in $\mu$-capture in hydrogen \cite{Bard} and in the recent
analysis of data on the $\pi^+$ electroproduction off the proton near the
threshold \cite{Choi}, are depicted. We see that their results agree with the
pion-pole dominance hypothesis in a large range of transfers, unlike our
result where this hypothesis is valid in a narrow $t$-range, and outside the
range the contribution of continuum is considerable. Note that the
contributions of the radial excitations of pion ($\pi^\p(1300)$ and
$\pi^\p(1770)$), which are rather distant from this region, are suppressed,
and their account would only slightly increase the mass of $\pi^\p(500)$). The
parameters of this pole term in \rf{G_P} might be changed more considerably if
the scalar $\sigma(555)$ and/or $\epsilon(750)$, discussed at present
\cite{Ishida}, are confirmed. Then it would be necessary to consider the
channel $(\epsilon\pi)$ and a possible multichannel nature of this state. At
all events, the conclusion about the necessity of the state in the range
500-800 MeV with $I^G(J^P)=1^-(0^-)$ for explaining the obtained IPE data will
remain valid. Note that recently the state with those parameters has been
observed in the $\pi^+\pi^-\pi^-$ system \cite{Ivan} and interpreted as the
first radial excitation of pion in the framework of a covariant formalism for
two-particle equations used for constructing a relativistic quark model
\cite{Skachkov}. Accepting this designation for $\pi^\p(500-800)$
and taking an estimation for the $\pi^\p$ weak-decay constant in the Nambu --
Jona-Lasinio model, generalized by using effective quark interactions with a
finite range, $f_{\pi^\p}=0.65$ MeV, we obtain $g_{\pi^\p N}=1.51$.
For this coupling constant, for now there are no suitable theoretical
calculations. In the NJL model, the consideration of radial excitations
of states
requires introducing some nonlocality. Since the successful calculation
of the $\pi N$ coupling constant in that model enforces one to go beyond the
framework of the tree approximation and take loop corrections into account
\cite{Nagy}, it seems that a satisfactory evaluation is possible in that
approximation with some nonlocality involved. Of course, more reliable
interpretation of $\pi^{\prime}$
would require investigation of other processes with $\pi^{\prime}$, and the
presence of this state would raise the question on its SU(3)-partners and on
careful (re)analyses of the corresponding processes in this energy region.

\section{Conclusion}

We see that a subsequent investigation of IPE is necessary for extracting
both a unique information about the electromagnetic structure of particles in
the sub-$N\overline{N}$ threshold region of the time-like momentum transfers
and the nucleon weak structure in the space-like region. The former is
especially interesting now, for example, in connection with the discussed
hidden strangeness of the nucleon (e.g. \cite{Gerasimov}) and quasinuclear
bound $p\bar p$ state \cite{Meshch}. Analyses of the experimental IPE data in
the first $\pi N$ resonance region allow one to obtain the $F_1^v$ values at
time-like transfers, which are quite consistent with the calculations in the
framework of the unitary analytic vector-meson dominance model \cite{Dubnicka}.
Furthermore, an inevitable step, necessary to study the electromagnetic
structure of nucleon-isobar systems in the time-like momentum-transfer region,
is a multipole analysis of IPE similar to that for photo- and
electroproduction (e.g. \cite{Dev-Lyth-Rank}). Setting experiments for
obtaining the data, aimed at carrying out that analysis, is possible at
present with intense pion beams being available.
When constructing the dispersion-quark model in the second and third
$\pi N$ resonance region, the multichannel character of the nucleon isobars
must be taken into account, e.g., with the help of the proper uniformizing
variables \cite{KMS-nc}.

It is relevant to mention a nuclear aspect of the above-described
analysis. As we have set in Sect.2, the $e^+ e^-$ production in collisions
of the 500~MeV/c positive pions with $^7$Li goes without disintegration of the
nucleus in about a half of events \cite{Baturin}. One can say that here first
one has observed the form factor of nucleus in the time-like momentum-transfer
region. However, when analyzing, the cross section on nucleus has been
supposed to be additively connected with the cross section on an individual
nucleon and a nuclear effect has been taken as screening into account. In
that analysis, unfortunately, a unique information on the electromagnetic
structure of the nucleus in the time-like region is lost. Generally, it seems
at present there is no satisfactory conception of the electromagnetic FF
of the nucleus in the time-like region. A satisfactory description must take
into account both a constituent character of the nucleus (and corresponding
analytic properties) and more subtle (than screening) collective nuclear
effects.

Finally, notice that a more reliable interpretation of the observed state
$\pi^{\prime}(500-800)$ requires to solve a number of questions, both
theoretical and experimental. In the pseudoscalar sector, states of various
nature are possible: except for $q\bar q$, the $gg$ and $ggg$ glueballs,
$q\bar q g$ hybrids, multiquark states. However, all the models and the
lattice calculations give masses of those unusual states, considerably greater
than 1 GeV; therefore, the most probable interpretation of
$\pi^{\prime}(500-800)$ does be the first radial pion excitation.

\section*{Acknowledgments}
The authors are grateful to S.B.Gerasimov, Yu.I.Ivanshin, V.A.Meshcheryakov,
L.L.Nemenov, G.B. Pontecorvo, N.B.Skachkov and M.K.Volkov for useful
discussions and interest in this work.

\underline{Figure Captions}\\

Fig.1:  Comparison of calculations in the CA approach to the
$\pi^-p\to\gamma^* n$ process for $E_{0+}-2E_{2-}$ with experimental data:
dashed and solid curves correspond to the cases when contribution to $G_P$ is
restricted by the pion pole $G_P^\pi$ and taken according to \rf{G_P},
respectively.\\

Fig.2:  The contributions to $G_P$ of possible intermediate states, coupled
with the current $A_\mu$: (a) one-pion state, (b) three-pion state,
(c) a resonance with the pion quantum numbers. \\

Fig.3:  The ratio of $G_P(t)/G_P^\pi(t)$. The curve corresponds to formula
\rf{G_P}. The points with errors (on the curve) indicate the error corridor
for this curve. The results of analysis of data on the $\mu$ capture in
hydrogen ($\bigtriangledown$) \cite{Bard} and on the $\pi^+$ electroproduction
off the proton near the threshold ($\bigtriangleup$) \cite{Choi} are depicted.


\newpage

\begin{figure}
\vskip 3.cm
\centerline{\epsfxsize=.7 \hsize \epsffile{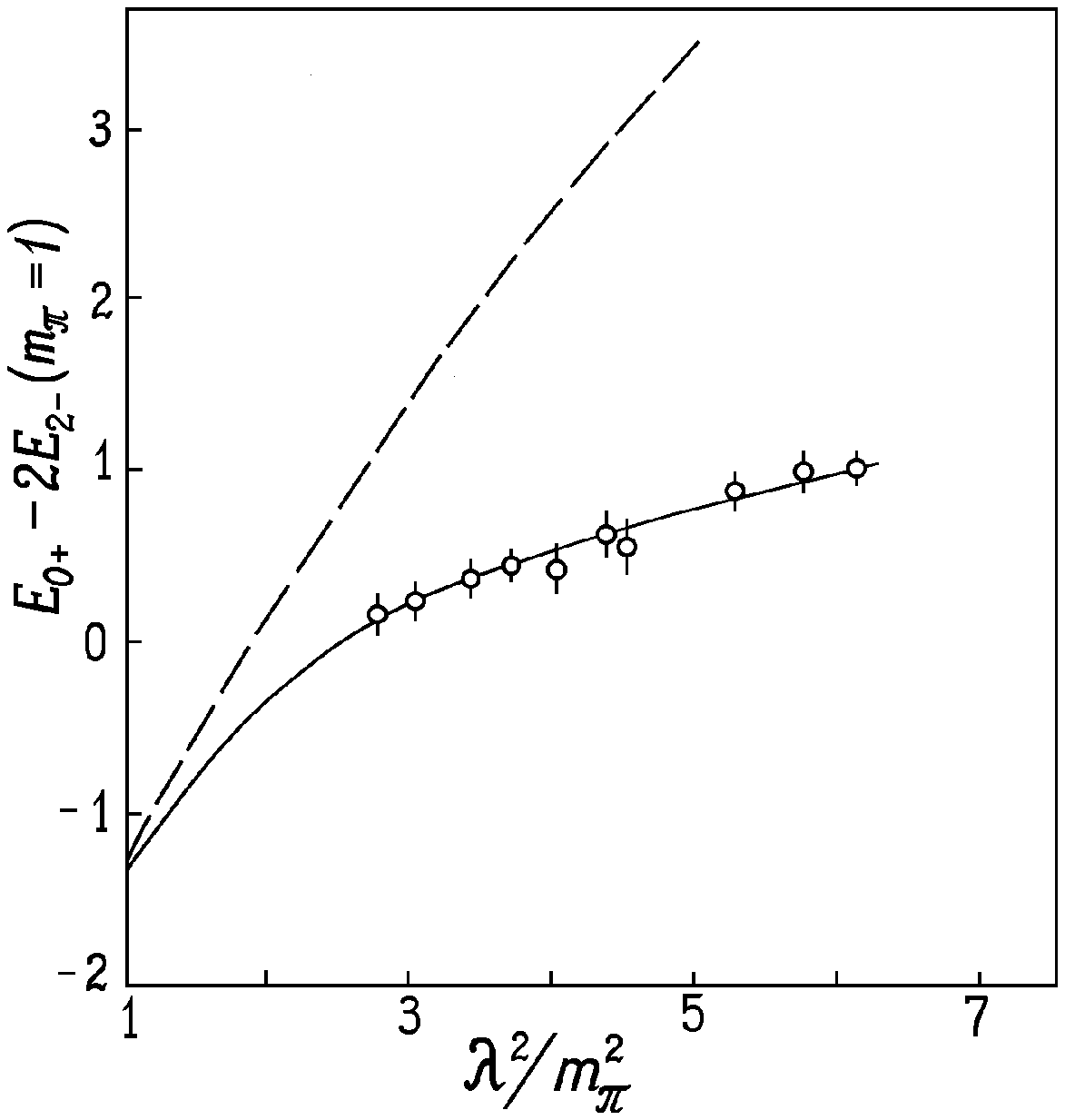}}
\vskip -.1cm
\caption{}
\end{figure}

\begin{figure}
\vskip 3.cm
\centerline{\epsfxsize=.7 \hsize \epsffile{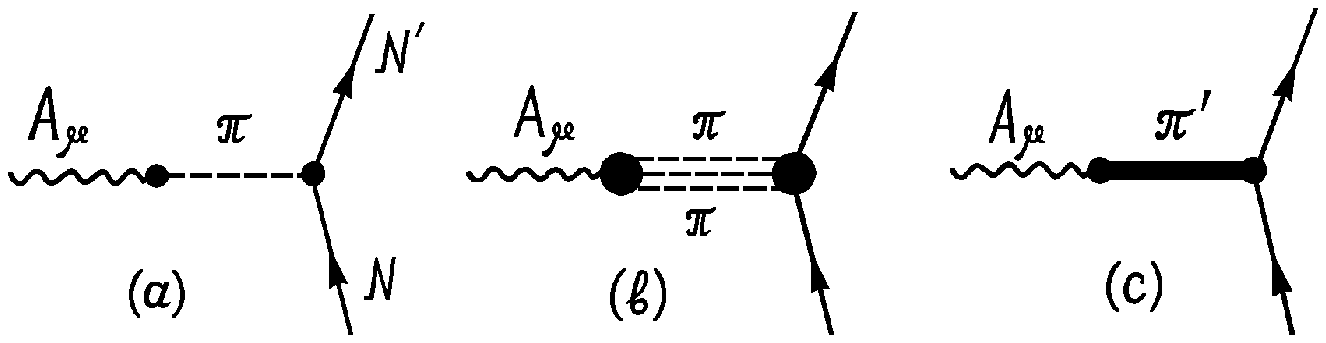}}
\vskip -.1cm
\caption{}
\end{figure}

\begin{figure}
\vskip 3.cm
\centerline{\epsfxsize=.7 \hsize \epsffile{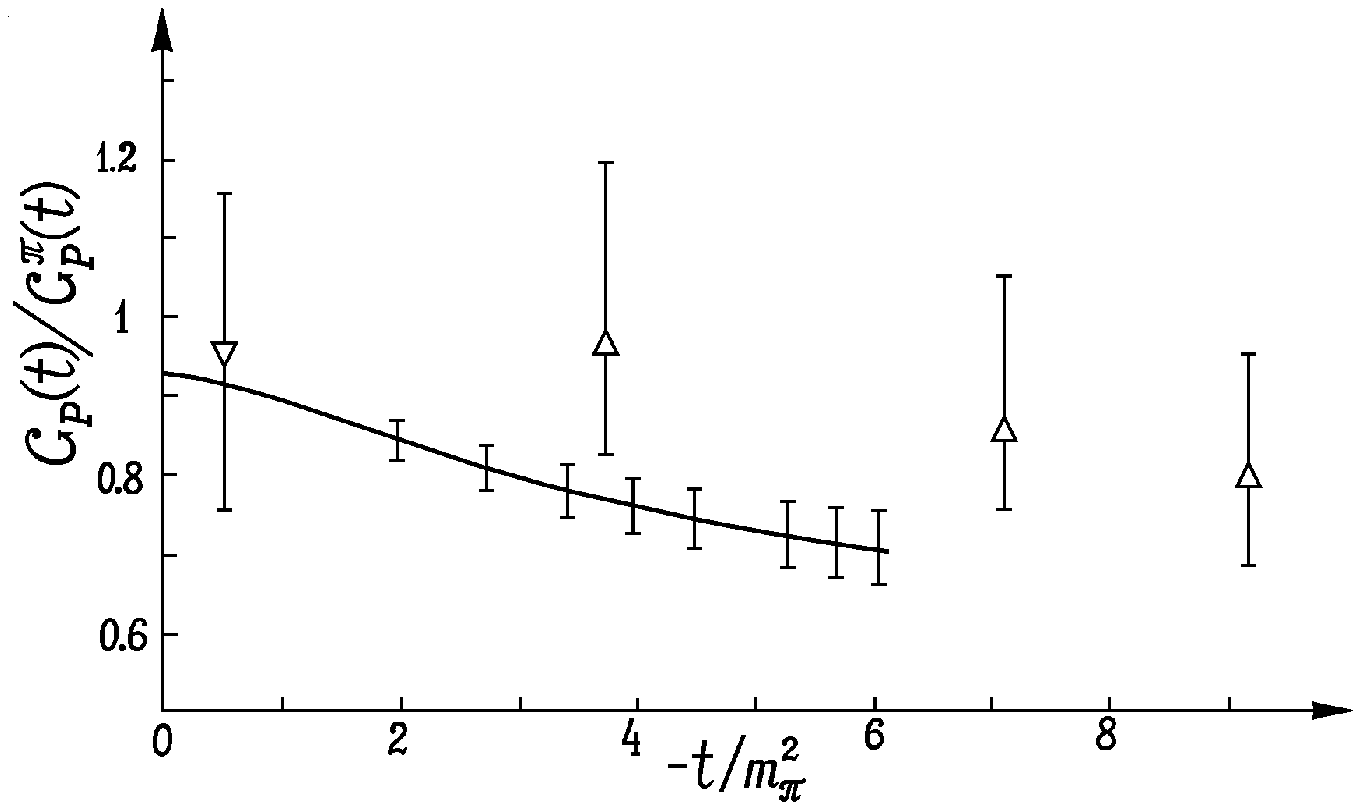}}
\vskip -.1cm
\caption{}
\end{figure}


\begin{thebibliography}{99}
\bibitem{Geffen} D.A.Geffen. -- Phys. Rev. {\bf 125} (1962) 1745.  J.Loubaton,
J.Tran Thanh Van. -- Nucl. Phys. {\bf B2} (1967) 342.
\bibitem{BST-yaf75} Yu.S.Surovtsev, F.G.Tkebuchava. -- JINR Communication
P2-4561, Dubna, 1969.  T.D.Blokhintseva, Yu.S.Surovtsev, F.G.Tkebuchava. --
Yad. Fiz. {\bf 21} (1975) 850.
\bibitem{Baldin} A.M.Baldin, V.A.Suleymanov. -- Phys. Lett. {\bf B37} (1971)
305;  JINR Communication P2-7096. Dubna, 1973.
 Yu.S.Surovtsev, F.G.Tkebuchava. -- Yad. Fiz. {\bf 55} (1992) 2138.
\bibitem{ST-yaf72} Yu.S.Surovtsev, F.G.Tkebuchava. -- Yad. Fiz. {\bf 16}
(1972) 1204.
\bibitem{Kulish} Yu.V.Kulish. -- Yad. Fiz. {\bf 16} (1972) 1102.
 G.Furlan, N.Paver, G.Verzegnassi. -- Nuovo Cim. {\bf A32} (1976) 75.
 N.Dombey, B.S.Read. -- J. Phys. G: Nucl. Phys. {\bf 3} (1977) 1659.
\bibitem{Tkeb} F.G.Tkebuchava -- Nuovo Cim. {\bf A47} {1978} 415.
\bibitem{Smirn-Shum} G.I.Smirnov, N.M.Shumeyko. -- Yad. Fiz. {\bf 17} (1973)
1266.  A.Bietti, S.Petrarca. -- Nuovo Cim. {\bf A22} (1974) 595; Lett. Nuovo
Cim. {\bf 13} (1975) 539.
\bibitem{ST-yaf75} Yu.S.Surovtsev, F.G.Tkebuchava. -- Yad.Fiz. {\bf 21} (1975)
753.
\bibitem{Samios} N.P.Samios. -- Phys. Rev. {\bf 121} (1961) 275.
 H.Kobrak. -- Nuovo Cim. {\bf 20} (1961) 1115.  S.Devons et al. -- Phys. Rev.
{\bf 184} (1969) 1356.  M.N.Khachaturyan et al. -- Phys. Lett. {\bf B24}
(1967) 349.  C.M.Hoffman et al. -- Phys. Rev. {\bf D28} (1983) 660.
H.Fonvieille et al. -- Phys. Lett. {\bf B233} (1989) 60, 65.
\bibitem{Berezh} Yu.K.Akimov et al. -- Yad. Fiz. {\bf 13} (1971) 748.
 S.F.Berezhnev et al. --Yad. Fiz. {\bf 16} (1972) 185; {\bf 18} (1973) 102;
{\bf 24} (1976) 1127; {\bf 26} (1977) 547.  G.D.Alekseev et al. -- Yad. Fiz.
{\bf 36} (1982) 322.  V.V.Alizade et al. -- Yad. Fiz. {\bf 33} (1981) 357;
{\bf 46} (1987) 1360.
\bibitem{Baturin} V.M.Baturin et al. -- Yad. Fiz. {\bf 47} (1988) 708.
\bibitem{Bardin} G.Bardin. -- Phys. Lett. {\bf B255} (1991) 149; {\bf B257}
(1991) 514.
\bibitem{Dev-Lyth-Rank} R.C.E.Devenish et al. -- Phys. Lett. {\bf B52} (1974)
227.  R.C.E.Devenish et al. -- Nuovo Cim. {\bf A1} (1971) 475.
\bibitem{Goghilidze} S.A.Goghilidze, Yu.S.Surovtsev, F.G.Tkebuchava. --
Yad. Fiz. {\bf 45} (1987) 1085.
\bibitem{Adler} S.L.Adler -- Ann. Phys. {\bf 50} (1968) 189.
\bibitem{Sur} Yu.S.Surovtsev -- Author's summary, 2-11047, Dubna, 1977.
\bibitem{Hohler} G.H\"ohler, E.Pietarinen -- Nucl. Phys. {\bf B95} (1975) 210.
\bibitem{Avakov} G.V.Avakov et al. -- Yad. Fiz. {\bf 44} (1986) 1471.
\bibitem{Dubnicka} S.Dubni{\'{c}}ka et al. -- Nuovo Cim. {\bf A106} (1993)
1253.
\bibitem{Volkov} M.K.Volkov, C.Weiss -- JINR Communication E2-96-131, Dubna,
1996.
\bibitem{Kalin} Yu.L.Kalinovsky, C.Weiss -- Z. Phys. {\bf C63} (1994) 275.
~I.V.Amirkhanov et al. -- ``Modelling'' {\bf 6} (1994) 57.
\bibitem{Bard} G.Bardin et al. -- Nucl. Phys. {\bf A352} (1981) 365;
Phys. Lett. {\bf B104} (1981) 320.
\bibitem{Brudanin} V.Brudanin et al. -- Nucl. Phys. {\bf A587} (1995) 577.
\bibitem{Choi} S.Choi et al. -- Phys. Rev. Lett. {\bf 71} (1993) 3927.
\bibitem{Ishida} Sh.Ishida et al. -- Progr. Theor. Phys. {\bf 95} (1996) 745;
hep-ph/9610359 v2 (27 May 1997). ~M.Svec -- Phys. Rev. {\bf D53} (1996) 2343.
\bibitem{Ivan} Yu.I.Ivanshin et al. -- Nuovo Cim. {\bf A107} (1994) 2855.
\bibitem{Skachkov} Yu.I.Ivanshin, N.B.Skachkov -- Nuovo Cim. {\bf A108} (1995)
1263.
\bibitem{Nagy} A.N. Ivanov, M. Nagy and N.I. Troitskaya: hep-ph/9805347.
\bibitem{Gerasimov} S.B.Gerasimov -- Chinese J. Phys. {\bf 34} (1996) 848.
\bibitem{Meshch} V.A.Meshcheryakov, G.V.Meshcheryakov -- Yad. Fiz. {\bf 60}
(1997) 1400.
\bibitem{KMS-nc} D.Krupa, V.A.Meshcheryakov, Yu.S.Surovtsev -- Nuovo Cim.
{\bf 109 A} (1996) 281.

\end{thebibliography}
\end{document}